\documentclass[twocolumn,preprintnumbers,amsmath,amssymb]{revtex4}
\usepackage{dcolumn}
\usepackage{bm}
\usepackage{graphicx,subfigure,xcolor}
\usepackage{braket}
\usepackage[normalem]{ulem}

\usepackage{comment}
\usepackage{xcolor}
\usepackage{amsmath}
\usepackage{amssymb}
\usepackage{eucal}
\usepackage{mathrsfs}
\usepackage{amsthm}
\usepackage{epstopdf}
\usepackage{textcomp}
\usepackage{braket}
\usepackage{pdfpages}

\begin{document}

\title{Resonance energy transfer between two atoms in a conducting cylindrical waveguide}

\author{Giuseppe Fiscelli}
\email{giuseppe.fiscelli@unipa.it}

\author{Lucia Rizzuto}
\email{lucia.rizzuto@unipa.it}

\author{Roberto Passante}
\email{roberto.passante@unipa.it}

\affiliation{Dipartimento di Fisica e Chimica, Universit\`{a} degli Studi di Palermo, Via Archirafi 36, I-90123 Palermo, Italy \\
and INFN, Laboratori Nazionali del Sud, I-95123 Catania, Italy}

\begin{abstract}
We consider the energy transfer process between two identical atoms placed inside a perfectly conducting cylindrical waveguide. We first introduce a general analytical expression of the energy transfer amplitude in terms of the electromagnetic Green's tensor; we then evaluate it in the case of a cylindrical waveguide made of a perfect conductor, for which analytical expressions of the Green's tensor exist. We numerically analyse the energy transfer amplitude when the radius of the waveguide is such that the transition frequency of both  atoms is below the lower cutoff frequency of the waveguide, so that the resonant photon exchange is strongly suppressed. We consider both cases of atomic dipoles parallel and orthogonal to the axis of the guide. In both cases, we find that the energy transfer is modified by the presence of the waveguide. In the near zone, that is when the atomic separation is smaller than the atomic transition wavelength, the change, with respect to the free-space case, is small for axial dipoles, while it is larger for radial dipoles; it grows when the intermediate region between near and far zone is approached. In the far zone, we find that the energy transfer amplitude is strongly suppressed by the waveguide, becoming virtually zero. A physical interpretation of these results is discussed. Finally, we discuss the resonance interaction energy and force between two identical correlated atoms in the waveguide, one excited and the other in the ground state, prepared in their symmetric or antisymmetric superposition.
\end{abstract}

\maketitle

\section{\label{sec:Introduction}Introduction}

Resonance energy transfer between quantum emitters is the exchange of excitation between them mediated by the quantum electromagnetic field \cite{JA99,CP98}. This process is of considerable importance in many different fields of physics, as well as in chemistry or biology, where coherent energy transfer between chromophores is supposed related to the very high efficiency in light-harvesting  observed in the photosynthesis process \cite{OCNF12,CS14}. It is also directly related to the resonance interaction force, that is a force resulting from the photon exchange between two atoms in the vacuum space, one excited and the other in the ground state, prepared in their symmetric or antisymmetric state \cite{CP98,NPR18}.
In the energy transfer process, the excitation, initially localised on one atom (donor), is transferred to the other atom (acceptor) through the electromagnetic field. Description of this process at different distances, i.e. near-field (radiationless), far-field (radiative or retarded) and the intermediate region where retardation effects start to appear, requires a full quantum-electrodynamical theory \cite{AS87,DJBA03}. For atoms in the free space, the energy-transfer amplitude behaves as $r^{-6}$ in the near-field region (F\"{o}rster limit) and as $r^{-2}$ in the far-field region \cite{AJ92}.

Since the pioneering work of Purcell, it is known that radiative processes of any quantum emitter(s), for example the spontaneous emission of one or more atoms, are affected by the environment \cite{Purcell46}. Radiation-mediated interactions, such as van der Waals and Casimir-Polder interactions, can be also significantly affected by the environment, that changes the photon density of states and the dispersion relation \cite{PT82,SPR06,fiscelli,haakh}, as well as from the presence of neighbouring atoms \cite{RPP07}, or due to a uniformly accelerated motion of the atoms \cite{NP13,MNP14}. Recently, investigations on how to control and tailor radiative processes through the environment have become a very active field of research, even in the case of time-modulated environments \cite{CRP17,Bagarello15,Antezza14}. For example, the effect of a structured environment such as a photonic crystal on the dipole-dipole interaction \cite{JW91,EGJ13,Hood16,Sproll18,SK13a}, or on the resonance interaction force between two entangled atoms \cite{Incardone14,NPR18,ZPR18}, has been investigated, showing possibility of enhancement or inhibition of the interaction. The effect of dispersive and absorbing surrounding media on the energy transfer between two atoms has been investigated \cite{DKW02,AF13}, as well as the possibility to control the resonance energy transfer between nanostructured emitters through a reflecting plate \cite{Weeraddana17,Jones18}.

In this paper we consider the resonance energy transfer between two identical atoms, or any other quantum emitter (for example quantum dots), placed on the axis of a cylindrical waveguide made of a perfect conductor. We first obtain an analytical expression of the energy transfer amplitude in terms of the electromagnetic Green's tensor of the cylindrical waveguide, whose expression is known. We then evaluate numerically the energy transfer amplitude in terms of the relevant parameters of the system, specifically the distance between the atoms and the radius of the waveguide, relative to the atomic transition wavelength. We consider both cases of atomic dipoles parallel and orthogonal to the guide axis. We explicitly show that the presence of a lower cutoff frequency inside the waveguide can deeply change the energy transfer amplitude, with respect to the case of atoms in the free space, in the far (radiative) zone, while in the near (radiationless) zone the change, although present, is much less important. A physical interpretation of these results is given. The relation of the results obtained with the resonance interaction energy and force between two identical entangled atoms inside the cylindrical waveguide is also discussed.

This paper is structured as follows. In Sec. \ref{sec:General formalism} we introduce our system and introduce a general expression of the resonance energy transfer between the atoms in terms of the electromagnetic Green's tensor of a generic environment. In Sec. \ref{Sec:3} we evaluate, using both analytical and numerical methods, the energy transfer amplitude for atoms inside a metallic cylindrical waveguide, showing how the guide can significantly change the energy-transfer process according to the relevant parameters of the system. In Sec. \ref{sec:4} we discuss the relevance of our results for the resonance interaction energy between correlated atoms. Finally, Sec. \ref{sec:Conclusions} is devoted to our conclusions and final remarks.

\section{\label{sec:General formalism} Energy transfer in terms of the electromagnetic Green's tensor}

We first introduce the energy transfer process between two identical two-levels atoms (or quantum dots), located inside a generic macroscopic structured environment and interacting with the quantum electromagnetic field.
Let us consider two atoms, labeled with A and B, and suppose
that atom A (donor) is in its excited state, while atom B (acceptor) is in its ground state. The atom A can decay and emit a real or virtual photon that can be absorbed by atom B. Excitation is thus transferred from donor to acceptor atom through the electromagnetic field \cite{andrews1}. In this section we obtain the energy transfer rate between the two atoms in a generic macroscopic environment, whose properties are described by its electromagnetic Green's tensor \cite{Gruner96,DKW02,Buhmann1}.
According to the generalized Fermi golden rule \cite{JA99,AJ92},  this is given by
\begin{equation}
W_{i\rightarrow f}= \frac{2\pi}{\hbar}|\bra{\psi_i}T\ket{\psi_f}|^2 \delta(E_f-E_i) ,
\end{equation}
where $ \ket{\psi_i}=\ket{e_A,g_B,0} $ and $\ket{\psi_f}=\ket{g_A,e_B,0}  $ are respectively the initial and the final state of the two-atom system, with energy $E_i$ and $E_f$. $\mid g_{A/B} \rangle$ and $\mid e_{A/B}\rangle$ are respectively the ground and excited atomic states, and $\mid 0 \rangle$ represents the photon vacuum state in the presence of the external environment. We assume the two atoms identical, with transition frequency $\omega_0 = c k_0$. $T$ is the transition operator at the second order,
\begin{equation}
T=H_i+H_i\frac{1}{E_i-H_0}H_i ,
\end{equation}
where $H_0$ and $H_i$ are the unperturbed and interaction Hamiltonians, respectively, and $E_i=E_f$ is the energy of the initial and final states. The  second-order energy-transfer amplitude between the two atoms can be written as
\begin{equation}
M=\bra{\psi_i}T\ket{\psi_f}=\sum_I \frac{\bra{\psi_i}H_i\ket{I}\bra{I}H_i\ket{\psi_f}}{E_i-E_I} ,
\label{ET-amplitude}
\end{equation}
where $ \ket{I} $ are the intermediate states with energy $E_I$, that can contribute to the energy transfer process.
If both atoms are in the free space, the energy transfer amplitude is given by the well-known expression \cite{AS87,JDA04,Salam08,Salam10}
\begin{eqnarray}
M_{fs}^{(\pm )} &=& \frac{d_{Ai}^{ge}d_{Bj}^{eg}}{4\pi\epsilon_0r^3} \Bigg[   \big(\delta_{ij} - 3\mathbf{\hat{r}}_i\mathbf{\hat{r}}_j\big) \left( 1 \pm ik_0 r \right) \nonumber  \\
&-&\big(\delta_{ij} - \mathbf{\hat{r}}_i\mathbf{\hat{r}}_j\big)k_0^2r^2  \Bigg] e^{\mp ik_0 r} ,
\label{ET-vuoto}
\end{eqnarray}
where the $\pm$ sign refers to the two possible choices of circumventing the pole in $k=k_0$ by adding a small positive or negative imaginary part $\pm i\eta$ to the energy denominator in (\ref{ET-amplitude}), and we have used the Einstein notation for repeated indices. $k_0$ is the atomic transition wavenumber associated to its transition frequency, $r$ is the distance between the atoms and $d_{\alpha i}^{ge}= \bra{g}d_{\alpha i}\ket{e}$ ($\alpha=A, B$) their dipole-moment matrix elements. For a random orientation of the atomic dipole moments, Eq. (\ref{ET-vuoto}) yields a monotonic distance dependence of the energy transfer rate, proportional to \cite{JDA04,Salam10}
\begin{equation}
\mid M_{fs}^{(\pm )} \mid^2 = \frac {2 \mid \mathbf{d}_A\mid^2  \mid \mathbf{d}_B \mid^2}{(4\pi\epsilon_0r^3)^2}\left( 3 +k_0^2 r^2 + k_0^4 r^4 \right) .
\label{energytransferrate}
\end{equation}

We now assume the atoms placed in a generic linear magneto-dielectric environment. It is well known that the presence of the environment can significantly affect the resonant energy transfer process \cite{DKW02,Weeraddana17, Salam12}.
To investigate the excitation exchange between the two atoms inside a structured environment, we exploit a procedure based on the Green's tensor formalism \cite{Gruner96,DKW02,Buhmann1,BW07}.
This method has been widely used in many contexts, from quantum electrodynamics to quantum optics, and its merit is that all relevant properties and effects of the environment are included in the Green's tensor expression. It has been used, for example, to evaluate van der Waals and Casimir-Polder forces in external environments \cite{SBWD06,EBS10,SK13,haakh,BPRB16}, or to investigate the collective spontaneous decay of two quantum emitters placed nearby a reflecting mirror \cite{Palacino17}.
We will use this approach to obtain the energy transfer amplitude when two emitters are inside an environment such as a conducting cylindrical waveguide. We first briefly review the method and the relevant expressions for the energy transfer in a generic linear magnetodielectric environment \cite{DKW02}. In the next section, we will then specialise our considerations to the specific case of a perfectly conducting cylindrical waveguide, which is the main point of this paper.

The Green's tensor $\mathbf{G}(\mathbf{r},\mathbf{r'},\omega)$ is defined as the solution of the Helmholtz equation (see, for example, \cite{Buhmann1,BW07})
\begin{equation}
\Biggl[\nabla\times \frac 1{\mu (\mathbf{r},\omega)}\nabla\times \, - \frac{\omega^2}{c^2}\epsilon(\mathbf{r},\omega)\Biggr]\mathbf{G}(\mathbf{r},\mathbf{r'},\omega)  = \mathbf{\delta}(\mathbf{r}-\mathbf{r'}) ,
\label{eqHelmotz}
\end{equation}
with the boundary condition $\mathbf{G}(\mathbf{r},\mathbf{r'},\omega)\rightarrow 0$ for $|\mathbf{r}-\mathbf{r'}|\rightarrow \infty $, and where $\epsilon(\mathbf{r},\omega)$ and $\mu (\mathbf{r},\omega)$ are respectively the electric and magnetic permittivity of the medium.
The medium-assisted electric field operator is expressed as
\begin{align}
\mathbf{E}(\mathbf{r})&=\int_0^\infty \! \! d\omega \mathbf{E}(\mathbf{r},\omega)+H.c. \notag\\
&=\sum_{\lambda=e,m}\int d^3r'\int_0^\infty \! \! d\omega \mathbf{G}_\lambda(\mathbf{r},\mathbf{r'},\omega)\cdot \mathbf{f}_\lambda(\mathbf{r'},\omega) +H.c. ,
\label{campoE-G}
\end{align}
where
\begin{eqnarray}
\mathbf{G}_e(\mathbf{r},\mathbf{r'},\omega)=i\frac{\omega^2}{c^2}\sqrt{\frac{\hbar}{\pi\epsilon_0}\mathrm{Im}\, \epsilon(\mathbf{r'},\omega)}\mathbf{G}(\mathbf{r},\mathbf{r'},\omega) \\
\mathbf{G}_m(\mathbf{r},\mathbf{r'},\omega)=i\frac{\omega}{c}\sqrt{-\frac{\hbar}{\pi\epsilon_0}\mathrm{Im} \, \frac 1{\mu (\mathbf{r'},\omega)}}[\nabla'\times\mathbf{G}(\mathbf{r},\mathbf{r'},\omega)]^T
\end{eqnarray}
are respectively the electric and magnetic Green's tensor components. They satisfy the following relation
\begin{eqnarray}
&& \sum_{\lambda=e,m}\int d^3s \mathbf{G}_\lambda(\mathbf{r},\mathbf{s},\omega)\cdot\mathbf{G}_\lambda^{*T}(\mathbf{r'},\mathbf{s},\omega)
\nonumber \\
&& \qquad = \frac{\hbar\mu_0}{\pi}\omega^2 \mathrm{Im} \mathbf{G}_\lambda(\mathbf{r},\mathbf{r'},\omega) \label{Proprietà green} .
\end{eqnarray}
The bosonic matter-assisted operators $\mathbf{f}_\lambda^\dagger(\mathbf{r},\omega) $ and  $ \mathbf{f}_\lambda(\mathbf{r},\omega) $ in \eqref{campoE-G} are respectively the creation and annihilation operators describing the combined system of the electromagnetic field and the magnetodielectric medium.
They satisfy the following commutation relations
\begin{align}
\Big[ f_{\lambda i}(\mathbf{r},\omega) , f_{\lambda' i'}^\dagger (\mathbf{r'},\omega')  \Big] &= \delta_{\lambda\lambda'}\delta_{ii'} \delta(\mathbf{r}-\mathbf{r'})\delta(\omega-\omega') , \notag \\
\Big[ f_{\lambda i}(\mathbf{r},\omega) , f_{\lambda' i'}(\mathbf{r'},\omega')  \Big] &= 0 ,
\label{commutatoref}
\end{align}
where the subscript $\lambda =e, m$ refers to the electric and magnetic parts.

The Hamiltonian of our system can be expressed as
\begin{equation}
H=H_a+H_f+H_i
\end{equation}
where $H_a$ and $H_f$ are  respectively the unperturbed atomic and field Hamiltonians (in the presence of the medium), given by
\begin{eqnarray}
H_a= \sum_{n=e,g} E_n^A \ket{n_A}\bra{n_A}+ \sum_{n=e,g} E_n^B \ket{n_B}\bra{n_B},\\
H_f= \sum_{\lambda=e,m} \int d^3r\int_0^{\infty}\! \! d\omega \, \hbar \omega \, \mathbf{f}_\lambda^\dagger(\mathbf{r},\omega)\cdot \mathbf{f}_\lambda(\mathbf{r},\omega) ,
\end{eqnarray}
where we have modeled the atoms as two-level systems ($|e \rangle$ and $| g \rangle$ being the excited and ground state with energy $E_e$ and $E_g$, respectively), and $ H_i $ is the interaction Hamiltonian in the multipolar coupling scheme, within the dipole approximation,
\begin{equation}
H_i=- \mathbf{d}_A\cdot\mathbf{E}(\mathbf{r}_A) - \mathbf{d}_B\cdot\mathbf{E}(\mathbf{r}_B)
\label{Hi} .
\end{equation}
Here $ \ket{n_{A(B)}} $ are eigenstates of the atomic Hamiltonian of atom A(B) at position $\mathbf{r}_{A(B)}$, $\mathbf{d}_{A(B)}$ is the atomic electric dipole moment operator, and $\mathbf{E}(\mathbf{r})$ is the electric field operator evaluated at the atomic position $\mathbf{r}=\mathbf{r}_{A(B)}$.

Using second-order perturbation theory, the energy transfer amplitude $M$ is (see Eq. \eqref{ET-amplitude})
\begin{equation}
M^{(\pm )}= \sum_I \frac{\bra{e_A,g_B,0}H_i\ket{I}\bra{I}H_i\ket{g_A,e_B,0}}{E_i-E_I \pm i\eta}
\label{deltaE} ,
\end{equation}
where $E_i = \hbar \omega_0$, with $ \omega_0 = ck_0$ the transition frequency of the atoms, and $\eta \rightarrow 0^+$. Taking into account Eq. \eqref{Hi}, only two intermediate states $\ket{I}$  contribute to the energy transfer amplitude \eqref{deltaE}
\begin{eqnarray}
\ket{I}_1= \ket{g_A,g_B,\mathbf{1}_\lambda(\mathbf{r},\omega)} , \notag \\
\ket{I}_2= \ket{e_A,e_B,\mathbf{1}_\lambda(\mathbf{r},\omega)} ,
\label{Intst}
\end{eqnarray}
where $\mathbf{1}_\lambda(\mathbf{r},\omega)$ is a medium-assisted excitation of the field.
The sum over the intermediate states \eqref{Intst} in Eq. (\ref{deltaE}) can be written as a sum over $\lambda$ (electric and magnetic field modes), and an integral over space and frequency, that is
\begin{multline}
M^{(\pm )} = \sum_{\lambda=e,m} \int d^3\mathbf{r} \int_0^\infty d\omega \\ \times\Bigg[ \frac{\bra{\psi_i}H_i\ket{g_A,g_B,\mathbf{1}_\lambda(\mathbf{r},\omega)}\bra{g_A,g_B,\mathbf{1}_\lambda(\mathbf{r},\omega)}H_i\ket{\psi_f}}{E_i-E_{I_1} \pm i\eta}  \\ +\frac{\bra{\psi_i}H_i\ket{e_A,e_B,\mathbf{1}_\lambda(\mathbf{r},\omega)}\bra{e_A,e_B,\mathbf{1}_\lambda(\mathbf{r},\omega)}H_i\ket{\psi_f}}{E_i-E_{I_2}}  \Bigg] \label{DeltaE}
\end{multline}
(the energy denominator in the second term of \eqref{DeltaE}  does not vanish in the integration range of $\omega$).

Using the expression (\ref{campoE-G}) of the electric field in terms of the Green's tensor, the interaction Hamiltonian (\ref{Hi}), commutation relations (\ref{commutatoref}) and the relation (\ref{Proprietà green}), after some algebra we have
\begin{multline}
M^{(\pm )}= \frac 1{\pi\epsilon_0c^2}\int_0^\infty d\omega \omega^2 \sum_{ij}\Bigg\{ \frac 1{\omega_0-\omega \pm i\eta}\\
\times \Bigg[  d_{Ai}^{eg} \mathrm{Im}G_{ij}(\mathbf{r}_A, \mathbf{r}_B,\omega)d_{Bj}^{ge} \Bigg] \\
 -  \frac 1{\omega_0+\omega} \Bigg[ d_{Ai}^{eg} \mathrm{Im}G_{ij}(\mathbf{r}_B, \mathbf{r}_A,\omega)d_{Bj}^{ge} \Bigg]   \Bigg\} \label{DeltaE-G} .
\end{multline}

Equation (\ref{DeltaE-G}) gives the amplitude probability that the electronic excitation is transferred from one atom to the other, when the atoms are placed inside a generic linear magnetodielectric environment, whose properties are expressed in terms of the electromagnetic Green's tensor. A prescription to avoid the resonant pole must be specified, similarly to the free-space case.

We wish to stress that the process discussed above is also directly
related to a quantum interaction energy between the atoms. It is the resonance interaction energy between two atoms, one in an excited state and the other in the ground state, prepared in a correlated (symmetric or antisymmetric) state  in the photon vacuum \cite{NPR18}. Indeed, the second-order energy shift $\Delta E$ due to the atom-field interaction (exchange of one real or virtual photon between the atoms) is given by
\begin{equation}
\Delta E_\pm= {\cal P}\sum_I \frac{\bra{\psi_\pm}H_i\ket{I}\bra{I}H_i\ket{\psi_\pm}}{\hbar \omega_0 -E_I} ,
\end{equation}
where ${\cal P}$ indicates the principal value, and  the state $\ket{\psi_\pm}$ is
\begin{equation}
\ket{\psi_\pm}=\frac{1}{\sqrt{2}}(\ket{e_A,g_B,0}\pm\ket{g_A,e_B,0}) ,
\label{entangedstates}
\end{equation}
that is the symmetric or antisymmetric entangled state.
In such a case, the excitation is delocalized between the atoms.

Following the same procedure used before, we obtain a general expression of the distance-dependent resonance interaction energy between the two atoms in terms of the Green's tensor of a generic environment (apart single-atom energy corrections that do not contribute to the interatomic force)
\begin{multline}
\Delta E_{\pm} = \pm\frac 1{2\pi\epsilon_0c^2} {\cal P} \int_0^\infty d\omega \omega^2 \sum_{ij}\Bigg\{\frac{1}{\omega_0-\omega} \\ \times\Bigg[  d_{Ai}^{eg} \mathrm{Im}G_{ij}(\mathbf{r}_A, \mathbf{r}_B,\omega)d_{Bj}^{ge}
 +  d_{Bi}^{eg} \mathrm{Im}G_{ij}(\mathbf{r}_B, \mathbf{r}_A,\omega)d_{Aj}^{ge}  \Bigg] \\
 -   \frac{1}{\omega_0+\omega} \Bigg[  d_{Bi}^{ge} \mathrm{Im}G_{ij}(\mathbf{r}_B, \mathbf{r}_A,\omega)d_{Aj}^{eg} \\
  + d_{Ai}^{ge} \mathrm{Im}G_{ij}(\mathbf{r}_A, \mathbf{r}_B,\omega)d_{Bj}^{eg}  \Bigg]   \Bigg\} .
\label{R-G}
\end{multline}
The resulting interatomic force between the two entangled atoms, prepared in their symmetrical (+) or anti-symmetrical (-) superposition, is then obtained by taking the derivative of \eqref{R-G} with respect to $r= \mid \mathbf{r}_A - \mathbf{r}_B \mid$, changed of sign (quasi-static approach).
In the last section of this paper we will specialise this result to the case of two atoms inside a perfectly conducting cylindrical waveguide.
Taking into account that $G_{ij}(\mathbf{r}, \mathbf{r}',\omega)= G_{ji}(\mathbf{r}', \mathbf{r},\omega)$ \cite{Buhmann1}, Eqs. (\ref{R-G}) and  (\ref{DeltaE-G}) show the strict relation between the resonance interaction energy $\Delta E_{\pm}$  and the resonance energy transfer amplitude $M$.
In the free space, Eq. (\ref{R-G}) yields an interaction energy scaling as $r^{-3}$ in the near (nonretarded) zone and as $r^{-1}$ in the far (retarded) zone, with space oscillations \cite{CP98,NPR18} (in the case of the energy shift, the principal-value prescription around the resonance pole yields space oscillations, contrarily to the monotonic result for the energy transfer where a different way to avoid the pole is used).

\section{Energy transfer between two atoms inside a conducting cylindrical waveguide}
\label{Sec:3}

\begin{figure}[h]
\centering
\includegraphics[width=0.3\textwidth]{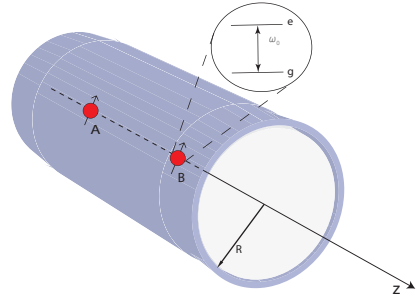}
\caption{The physical system: two atoms on the axis of a perfectly conducting cylindrical waveguide.}
\label{fig: sistema1}
\end{figure}

We now specialise our investigation to the case of two identical two-level atoms, one excited and the other in its ground state, placed inside a perfectly conducting cylindrical waveguide, as shown in Fig. \ref{fig: sistema1}.
The waveguide consists of a perfectly conducting cylindrical shell of radius $R$; we suppose that the atoms are located on the axis of the waveguide, $z$ being the interatomic distance. According to Eq. (\ref{DeltaE-G}), in order to obtain the energy transfer amplitude between the two atoms, we need the expression of the electromagnetic Green's tensor for the cylindrical waveguide.

The analytical expression of Green's function of a cylindrical waveguide, with the appropriate boundary conditions, is known in the literature and it has the following form \cite{tai} (see also \cite{haakh,Li00})
\begin{equation} \label{greencilindro1}
\begin{split}
&\mathbf{ G}(\mathbf{r},\mathbf{r'},\omega) = - \frac{1}{k^2}\delta(\mathbf{r}-\mathbf{r'}) \hat{z}\otimes\hat{z} \\&
+ \sum_{n,m} \Bigg[ c_{\mu n} \mathbf{M}_{^e_on\mu}(\pm k_\mu) \mathbf{M}_{^e_on\mu}'(\mp k_\mu) \\& + c_{\lambda n} \mathbf{N}_{^e_on\lambda}(\pm k_\lambda) \mathbf{N}_{^e_on\lambda}'(\mp k_\lambda)  \Bigg] \qquad (z\gtrless z') ,
\end{split}
\end{equation}
where the $ \mathbf{M} $ and $ \mathbf{N}$ vector cylindrical wave functions are given by
\begin{equation}
\mathbf{M}_{^e_om\mu}(h) = \Bigg[\mp \frac{nJ_n(\mu r)}{r}\begin{matrix} \sin \\ \cos \end{matrix} (n\phi)\hat{r}- \frac{\partial J_n(\mu r)}{\partial r} \begin{matrix} \cos \\ \sin \end{matrix} (n\phi) \hat{\phi}    \Bigg] e^{ihz} \label{eq M} ,
\end{equation}
\begin{equation}
\begin{split}
\mathbf{N}_{^e_on\lambda}(h) = \frac{1}{k} \Bigg[& ih \frac{\partial J_n(\lambda r)}{\partial r} \begin{matrix} \cos \\ \sin \end{matrix} (n\phi) \hat{r} \mp \frac{ihn}{r} J_n(\lambda r) \begin{matrix} \sin \\ \cos \end{matrix} (n\phi)\hat{\phi}  \\ &+ \lambda^2 J_n(\lambda r) \begin{matrix} \cos \\ \sin \end{matrix}  (n\phi) \hat{z}  \Bigg]e^{ihz} \label{eq N}.
\end{split}
\end{equation}
Here, $ \mu=q_{nm}/R $ and $ \lambda=p_{nm}/R $, with $ p_{nm} $ the \textit{m}th root of the $n$-order Bessel function ($ J_n(p_{nm})=0$), and $ q_{nm} $ the \textit{m}th root of the derivative of the $n$-order Bessel function ( $ J_n'(q_{nm})=0$). $ \lambda$  and $ \mu $ are the radial components of the wavevector $\mathbf{k}$ $(k=\frac{\omega}{c})$ of the electric field component for, respectively, the transverse magnetic (TM) and transverse electric (TE) modes inside the waveguide; likewise, $k_{\lambda} $ and $ k_{\mu}$ are the axial components of the wavevector. Since the atoms are placed on the cylinder's axis, the Green's tensor (\ref{greencilindro1}) can be simplified as \cite{haakh}
\begin{multline}
\mathbf{ G}(\mathbf{r},\mathbf{r'},\omega)=\frac{i}{4\pi} \sum_m \Bigg[ \Bigg(  \frac{e^{ik_\mu z}}{2I_{\mu 1}k_\mu} + \frac{k_\lambda e^{ik_\lambda z}}{2I_{\lambda 1}k^2} \Bigg) \\
\times(\hat{r}\otimes\hat{r} + \hat{\phi}\otimes\hat{\phi}) + \frac{\lambda^2e^{ik_\lambda z}}{I_{\lambda 0}k_\lambda k^2} \hat{z}\otimes\hat{z}  \Bigg] ,
\label{Green cilindrico}
\end{multline}
where
\begin{eqnarray}
I_{\mu 1}= \frac{R^2}{2}\Bigg(1-\frac{1}{q_{1m}^2}  \Bigg)  J_1^2(q_{1m}) , \\
I_{\lambda 1}= \frac{R^2}{4}\big(J_0(p_{1m})-J_2(p_{1m}) \big)^2 , \\
I_{\lambda 0}= \frac{R^2}{2}J_1^2(p_{0m}) ,
\end{eqnarray}
and
\begin{equation}
k_{\lambda}=\sqrt{k^2-\lambda^2} ,
\label{kl}
\end{equation}
\begin{equation}
k_{\mu}=\sqrt{k^2-\mu^2} .
\end{equation}

The presence of the cylindrical waveguide changes the density of states of the electromagnetic field inside it, and in particular determines a lower cut-off frequency for the TE and TM modes inside the waveguide, given by
\begin{equation}
(\omega_{min})_{TM} \simeq \frac{2.4c}{R} \quad ; \quad (\omega_{min})_{TE} \simeq \frac{1.8c}{R} .
\label{cutoff}
\end{equation}
Since the value of $ \omega_{min} $ depends from the waveguide radius $R$, it is possible to modify the electromagnetic field modes allowed inside the waveguide, and in particular its lower cutoff frequency, by modifying $R$. It is thus possible to control the energy transfer between the atoms, by changing $R$.

The presence of a lower cut-off frequency, as we will now show, has a strong effect on the excitation transfer between the atoms: if $k_0R\ll 1$, the atomic transition frequency is smaller than the waveguide cut-off frequency ($\omega_0<\omega_{min}$), and thus the waveguide suppresses the e.m. field modes resonant with the atomic transition frequency. Since they cannot contribute to the exchange of excitation between the atoms, the energy transfer will be strongly suppressed in this regime. Otherwise, when $ \omega_0>\omega_{min}$ the resonant field modes do contribute to the excitation transfer, and we expect that the energy transfer amplitude will be much less influenced by the presence of the waveguide.

In this paper we mainly focus on the first regime above mentioned, $\omega_0<\omega_{min}$, where the presence of the guide is expected to be relevant. In this case, the frequency $ \omega_{min} $ is the lower limit of the frequency integral in Eq. (\ref{DeltaE-G}). Since the waveguide Green's tensor (\ref{Green cilindrico}) is symmetric with respect to the exchange of the atomic positions, $\mathrm{Im}\, G_{ij}(\mathbf{r}_A, \mathbf{r}_B,\omega)= \mathrm{Im}\, G_{ij}(\mathbf{r}_B, \mathbf{r}_A,\omega)\,  (\forall i, j,  \omega )$, the energy transfer amplitude (\ref{DeltaE-G}) becomes
\begin{equation}
M =\frac{2}{\pi\epsilon_0} \sum_{ij}d_{Ai}^{eg}d_{Bj}^{ge}\int_{k_{min}}^\infty \! \! dk \frac{k^3}{ k_0^2-k^2} \mathrm{Im}\, G_{ij}(\mathbf{r}_A, \mathbf{r}_B,\omega) .
\label{equatione et}
\end{equation}
Since $ G_{ij}(\mathbf{r}_A, \mathbf{r}_B,\omega) $ is diagonal, we can write Eq. \eqref{equatione et} as a sum of three terms
\begin{align}
&M= M_z+M_r+M_{\phi},\\
&M_z=\frac{ ({\bf d}_{A}^{eg}\cdot\hat{z})({\bf d}_{B}^{ge}\cdot\hat{z})}{2\pi^2\epsilon_0}\sum_m \frac{\lambda^2}{I_{\lambda 0}}\int_{k_{min}}^\infty \! \!dk \frac{ k\cos(k_\lambda z)}{( k_0^2-k^2)k_\lambda}, \label{ETz}
\end{align}
\begin{align}
M_r=\frac{({\bf d}_{A}^{eg}\cdot\hat{r})({\bf d}_{B}^{ge}\cdot\hat{r})}{4\pi^2\epsilon_0}\sum_m \int_{k_{min}}^\infty \! \! dk \frac{ k^3}{ k_0^2-k^2} \notag\\
 \times \Bigg(  \frac{\cos (k_\mu z)}{I_{\mu 1}k_\mu} +\frac{k_\lambda \cos (k_\lambda z)}{I_{\lambda 1}k^2} \Bigg)\label{ETr},\\
M_{\phi}=\frac{({\bf d}_{A}^{eg}\cdot\hat{\phi})({\bf d}_{B}^{ge}\cdot\hat{\phi})}{4\pi^2\epsilon_0}\sum_m \int_{k_{min}}^\infty \! \! dk \frac{k^3}{ k_0^2-k^2}
\notag\\
 \times \Bigg(  \frac{\cos (k_\mu z)}{I_{\mu 1}k_\mu}  +\frac{k_\lambda \cos (k_\lambda z)}{I_{\lambda 1}k^2} \Bigg),
\label{ETphi}
\end{align}
where $k_{min}=\omega_{min}/c$.
As mentioned, we now investigate in detail the behavior of the energy transfer amplitude in the regime $Rk_0\ll 1$, as a function of the relevant parameters of the system: the interatomic distance $z=\mid\mathbf{r}_A- \mathbf{r}_B\mid$, the waveguide lower cut-off frequency $ \omega_{min} $ and the orientation of the atomic dipoles relative to the waveguide axis.
For symmetry reasons, we need to consider only the two following cases:
axial dipoles ($ M_z $ contribution) and radial dipoles ($ M_r $ contribution), relative to the waveguide axis.

\subsection{Axial dipoles}
We first assume that the atomic dipole moments are parallel and oriented along the positive $z$ axis (that is along the waveguide axis). In this case the energy transfer amplitude $M$ is given only by the $ M_z $ term, because \eqref{ETr} and \eqref{ETphi} vanish. Taking into account that $Rk_0 \ll 1$, the integral over $k$ in \eqref{ETz} becomes
\begin{equation}
\int_{k_{min}}^\infty \! \! dk \frac{ k\cos (k_\lambda z)}{( k_0^2-k^2)k_\lambda} =
 \int_0^\infty \! \! dk_\lambda \frac{\cos (k_\lambda z) }{k_0^2-\lambda^2 -k_\lambda^2}  ,
\label{int}
\end{equation}
where a change of variable has been done by using \eqref{kl}  ($\lambda$ is constant with respect to the integration variable $k_\lambda$). We wish to point out that in our case there is not a resonant pole at $k=k_0$ in the $k$ integrals, because $k_0 < k_{min}$.

The integral in (\ref{int}) has poles at the imaginary values $k_\lambda= \pm i \sqrt{\lambda^2 -k_0^2}$, and application of the residue theorem to the integral over $k_\lambda$ in (\ref{int}) yields
\begin{equation}
\frac{\pi e^{-\sqrt{\lambda^2-k_0^2}z}}{2\sqrt{\lambda^2-k_0^2}} .
\end{equation}

We thus obtain
\begin{equation}
M_{z}=\frac{ d_{Az}^{eg}d_{Bz}^{ge}}{2\pi\epsilon_0}\sum_m \frac{\lambda^2}{I_{\lambda 0}}\frac{e^{-\sqrt{\lambda^2-k_0^2}z}}{2\sqrt{\lambda^2-k_0^2}} .
\label{ETz2}
\end{equation}

The sum on $m$ in (\ref{ETz2}) is over all radial field modes allowed in the waveguide. Using the root test, it is possible to show that this series converges. We have evaluated numerically this quantity and verified explicitly that a good estimate for the energy transfer amplitude can be obtained by taking the first thirty terms of this sum (for the ranges we are considering).

We have evaluated numerically the energy transfer amplitude (\ref{ETz2}) as a function of the interatomic distance $z$, in two different regimes: $ z<\lambda_0 $ (near zone) and $ z>\lambda_0 $ (far zone). In our numerical evaluation, we have chosen the waveguide radius equal to $R=10^{-8}$m, and the atomic transition wavelength  $\lambda_0=5\cdot 10^{-7}$m.
\begin{figure}[h]
\centering
\includegraphics[width=0.52\textwidth]{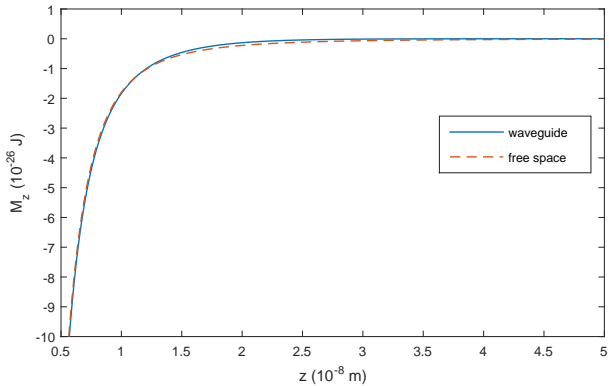}
\caption{Comparison between the energy transfer amplitude $M=M_z$ (axial dipoles) in the free space (orange dashed line) and in the waveguide (blue solid line), as a function of the interatomic distance $ z $, for $ z<\lambda_0$ (near-zone). The numerical values of the parameters are chosen such that $ \lambda_0=5\cdot 10^{-7}$ m, $ R=10^{-8}$ m, and $d_{A/B \, z}^{eg} =10^{-30}$ C $\cdot$ m.}
\label{fig:confrontointerazionez-near zone}
\end{figure}
\begin{figure}[h]
\centering
\includegraphics[width=0.52\textwidth]{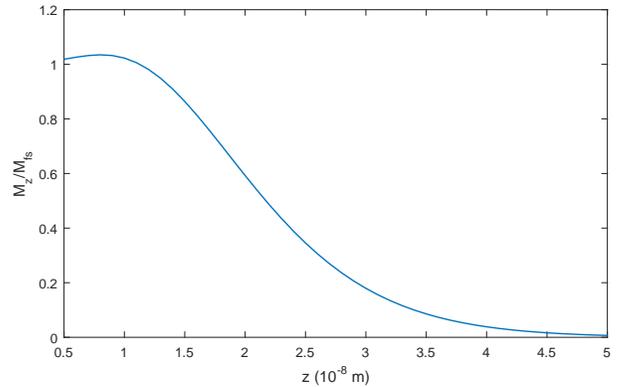}
\caption{Energy-transfer amplitude in the near zone between atoms with axial dipoles, normalised to the free-space energy-transfer amplitude, as a function of the interatomic distance $z$. When $z$ approaches the transition region between near and far zone ($z \sim \lambda_0$), the amplitude in the waveguide becomes more and more suppressed with respect to the free-space case. The parameters used are $ \lambda_0=5\cdot 10^{-7}$ m, $ R=10^{-8}$ m and $d_{A/B \, z}^{eg} =10^{-30}$ C $\cdot$ m.}
\label{fig:ET_rapporto_z}
\end{figure}

Fig. \ref{fig:confrontointerazionez-near zone} shows the energy transfer amplitude between the two atoms in the near zone as a function of the interatomic distance $z$, in the waveguide (blue continuous curve) and in the free space (orange dashed curve). The two plots show that the behavior with the distance is similar in the two cases. The plot in Fig. \ref{fig:ET_rapporto_z} represents the ratio between the energy transfer amplitude in the waveguide and in free space. It shows that for
$z \lesssim 1.2 \cdot 10^{-8}$m the two amplitudes are essentially the same, while for $z\gtrsim 1.2 \cdot 10^{-8}$m, when the intermediate region between the near and far zone is approached, the amplitude in the waveguide becomes more and more suppressed.

This result is related to the fact that in the near zone the interaction is essentially the electrostatic dipole-dipole interaction, which is not significantly modified by the waveguide. Approaching the intermediate region, $z \sim \lambda_0$, the effect of the waveguide on the amplitude becomes more relevant.
On the contrary, in the far zone, a numerical analysis shows that the energy transfer in the waveguide is strongly inhibited with respect to the free-space case by several orders of magnitude: it is virtually zero for $z>\lambda_0 $.

The results obtained show that, since in the regime $Rk_0\ll 1$ ($\omega_0<\omega_{min}$) the resonant
field modes are suppressed, the energy transfer in the waveguide is slightly modified in the very near zone, more and more suppressed approaching the intermediate region, and totally inhibited (virtually zero) in the far zone. The completely different effect of the waveguide in the near and far zones is due to the fact that in the near zone the process is essentially radiationless, while in the far zone it is a resonant radiative process, and that, in the cases considered, photons resonant with the atomic transition frequency cannot propagate in the guide.

\begin{figure}[h]
\centering
\includegraphics[width=0.52\textwidth]{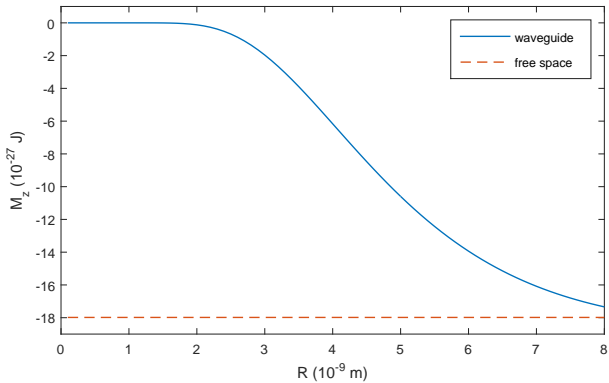}
\caption{The energy transfer amplitude $M_z$  for axial dipoles in the near zone, $z<\lambda_0$, between two atoms placed inside the cylindrical waveguide (blue solid line), as a function of the waveguide radius $R$. The orange horizontal line shows the value of the energy transfer amplitude in the free space. The numerical values of the parameters are $z=10^{-8}$ m, $\lambda_0=5\cdot 10^{-7}$ m and $d_{A/B \, z}^{eg} =10^{-30}$ C $\cdot$ m.}
\label{fig:raggioz-near zone}
\end{figure}

We can also investigate the energy transfer amplitude as a function of the waveguide cut-off frequency $ \omega_{min}\propto \frac{1}{R} $. By decreasing $R$, the cut-off frequency $ \omega_{min}$ increases and the gap between $ \omega_0 $ and $ \omega_{min} $ increases too, further reducing the energy-transfer process.
Fig. \ref{fig:raggioz-near zone} shows the numerical results obtained in the near zone for the energy-transfer amplitude, as a function of the waveguide radius $R$. The transition wavelength is $\lambda_0=5\cdot 10^{-7} $ m and the distance between the atoms is $z=10^{-8}$ m.
In this regime, when the waveguide radius $R$ is increased and thus the gap between $\omega_0$ and $\omega_{min}$ decreases, the absolute value of the energy transfer amplitude first increases and then settles to an almost constant value. On the contrary, by decreasing $R$, the energy transfer amplitude quickly tends to vanish, as expected.
These results show that the energy transfer between the atoms inside the waveguide can be strongly modified through the waveguide's radius $R$, both in the near and in the far zone.

\subsection{Radial dipoles}
\begin{figure}[h]
\centering
\includegraphics[width=0.52\textwidth]{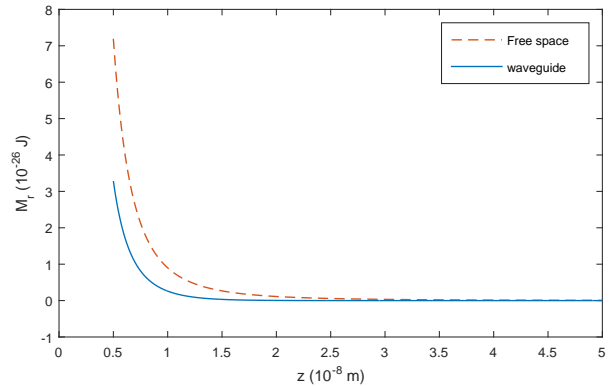}
\caption{The energy transfer amplitude for radial dipoles in the near zone, $ z<\lambda_0 $, as a function of the interatomic distance $ z $. The blue solid line is for atoms in the waveguide, while the orange dashed line refers to the free-space case. The parameters used are $ \lambda_0=5\cdot 10^{-7}$ m, $ R=10^{-8}$ m and $d_{A/B \, r}^{eg} =10^{-30}$ C $\cdot$ m.}
\label{fig:confrontointerazioner-near zone}
\end{figure}

We now consider the case of atomic dipole moments along the radial direction (that is orthogonal to the guide axis), and parallel to each other.
In this case the energy transfer amplitude $M$ is given only by the term $ M_r $, while the contributions $M_z$ and $M_{\phi}$ vanish (Eqs. (\ref{ETz}-\ref{ETphi})). The analytic expression of the energy transfer amplitude for radial dipoles, when $ R k_0\ll 1 $, can be obtained from Eq. (\ref{ETr}) which, after performing the integral over $k$, yields
\begin{multline}
M_{r}=\frac{ d_{Ar}^{eg}d_{Br}^{ge}}{8\pi\epsilon_0}\sum_m \Bigg(  -\frac{ k_0^2 e^{-\sqrt{\mu^2-k_0^2}z}}{I_{\mu 1}\sqrt{\mu^2-k_0^2}} \\+ \frac{ \sqrt{\lambda^2-k_0^2} e^{-\sqrt{\lambda^2-k_0^2}z}}{I_{\lambda 1}} \Bigg) .
\label{interazione rr}
\end{multline}
The root test shows that the sum over the radial field modes in (\ref{interazione rr}) converges, and a numeric check shows that the first forty terms of the sum give a good numerical estimate of it in the range of parameters we are considering.
As in the previous case of axial dipoles, we now investigate $M_{r}$ as a function of the interatomic distance $z$ and the waveguide cut-off frequency
$\omega_{min}$.
Fig. \ref{fig:confrontointerazioner-near zone} shows
the energy transfer amplitude in the near zone as a function of the interatomic distance $z$, using the same values as before for the atomic transition wavelength ($ \lambda_0=5\cdot 10^{-7}$ m) and the waveguide radius  ($R=10^{-8}$ m).
In this regime, the energy transfer amplitude inside the waveguide is reduced with respect to the free-space case, and this effect is much larger with respect of the previous case of axial dipoles. The reduction of the energy transfer amplitude significantly grows as the transition region between near and far zone is approached (for example, for $z=5\cdot 10^{-8}$m it is reduced of about three orders of magnitude).

In the far zone, our numerical analysis shows that the behaviour of the energy transfer is very similar to that for axial dipoles case: the waveguide strongly suppresses the energy transfer amplitude, which is virtually zero.
These results confirm that the energy transfer amplitude is significantly affected by the presence of the waveguide. The amplitude is more and more reduced, with respect to the free-space case, when the transition region at $z\sim \lambda_0$ is approached, and completely suppressed in the far zone.

\begin{figure}[h]
\centering
\includegraphics[width=0.52\textwidth]{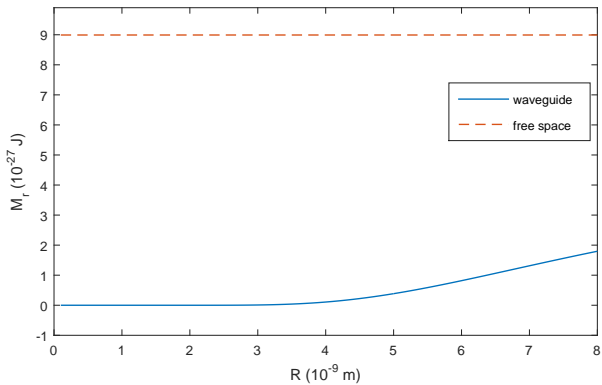}
\caption{The energy transfer amplitude $M_r$ for radial dipoles in the near zone ($z<\lambda_0$), as a function of the waveguide radius $R$, for atoms inside a cylindrical waveguide (blue continuous line). The orange horizontal line gives the value of the amplitude in the free space.. Parameters are $z=10^{-8}$m, $\lambda_0=5\cdot 10^{-7}$m and
$d_{A/B \, r}^{eg} =10^{-30}$ C $\cdot$ m.}
\label{fig:raggior-near zone}
\end{figure}

We now consider the excitation exchange between the atoms in the near zone, as a function of the waveguide cut-off frequency $\omega_{min}$ or equivalently of the waveguide radius $R$.
Fig. \ref{fig:raggior-near zone} shows that the energy transfer amplitude decreases for decreasing $R$, as expected because the difference between the atomic transition frequency $\omega_0$ and the waveguide cut-off frequency $\omega_{min}$, increases. In the range considered, it is smaller than in the free space, even if of the same order of magnitude.
\newline

In conclusion, our findings show that, for both axial and radial dipole orientations,  the presence of the waveguide allows us to control (reduce or inhibit, for example) the energy transfer process.

\section{Resonance dipole-dipole interaction energy between atoms inside the perfectly conducting waveguide}
\label{sec:4}

In this section we discuss the resonance dipole-dipole interaction energy between two entangled atoms placed on the axis of a perfectly conducting cylindrical waveguide, which is given by Eq. \eqref{R-G}. We consider the case $Rk_0\ll 1$, that is $\omega_0<\omega_{min}$. In this case the field modes resonant with the atomic transition frequency are suppressed by the waveguide, and do not contribute to the resonance interaction energy. Since the electromagnetic Green's tensor (\ref{Green cilindrico}) is symmetric for exchange of the positions of the atoms, Eq.  (\ref{R-G}) yields
\begin{equation}
\Delta E_\pm = \pm \frac{2}{\pi\epsilon_0} \sum_{ij}d_{Ai}^{eg}d_{Bj}^{ge}{\cal P}\int_{k_{min}}^\infty\! \! \frac{dk  k^3}{ k_0^2-k^2} \mathrm{Im}G_{ij}(\mathbf{r}_A, \mathbf{r}_B,\omega) ,
\label{equation R}
\end{equation}
where $+$ and $-$ signs refer, respectively, to the symmetric or antisymmetric entangled state in Eq. \eqref{entangedstates}.

Expression \eqref{equation R} coincides in modulus with the energy transfer amplitude (\ref{equatione et}). In fact, it should be noted that, due to our assumption $k_0 < k_{min}$, the resonant pole at $k=k_0$ is absent: thus, the principal value in (\ref{R-G}), as well as the $\pm i\eta$ prescription in (\ref{DeltaE-G}), do not play any role. Thus all results obtained in the previous section for the energy transfer amplitude, when the atomic transition frequency is below the waveguide lower cutoff frequency, can be easily extended to the resonance interaction energy between correlated atoms.  We stress that the interatomic resonance interaction and the resonant energy transfer process, are different albeit related physical processes: the resonance interaction (\ref{equation R}) is a quantum interaction energy between two atoms, one excited and the other in the ground state, prepared in a symmetrical or anti-symmetrical entangled state, arising from the exchange of a photon between them; it eventually yields a force between the atoms (in a quasi-static approach, the force is obtained from the derivative of the interaction energy with respect to the interatomic distance, changed of sign).
We discuss this interaction energy as a function of the interatomic distance $z$ and the waveguide cut-off frequency $\omega_{min}$.

The plot in Fig. \ref{fig:confrontointerazionez-near zone} shows that, in the near zone, the interaction between the two atoms (symmetric state) with axial dipole moments has essentially the same behaviour as in the free space. Approaching the transition region between near and far zone, the resonance interaction becomes more and more suppressed with respect to the free-space interaction, as it is highlighted in Fig. \ref{fig:ET_rapporto_z} where their ratio is plotted.

\begin{figure}[h]
\centering
\includegraphics[width=0.52\textwidth]{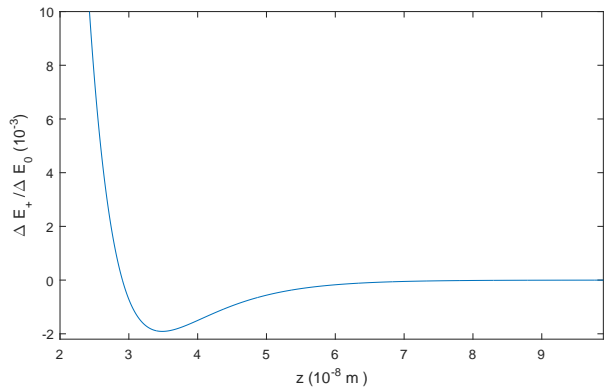}
\caption{Resonance interaction energy for the symmetric state in the near zone between atoms with radial dipoles, as a function of the interatomic distance $ z $, normalised to the free space interaction. When $z$ approaches the transition region between near and far zone ($z \sim \lambda_0$), the resonance interaction becomes more and more suppressed with respect to the free-space interaction. The parameters used are $ \lambda_0=5\cdot 10^{-7}$ m, $ R=10^{-8}$ m and $d_{A/B \, r}^{eg} =10^{-30}$ C $\cdot$ m.}
\label{fig:rapporto}
\end{figure}

More significant differences between the waveguide and free-space cases, emerge when we consider the radial dipole moments, in the near-zone ($ z<\lambda_0 $), as represented in Fig. \ref{fig:confrontointerazioner-near zone} for the symmetric state.
Now, the presence of the waveguide more deeply influences the resonance interaction between the atoms.
Fig. \ref{fig:rapporto} shows the ratio between the interaction energy in the waveguide and in the free space; it highlights that the resonance interaction energy changes from repulsive to attractive, for distances greater than $z\gtrsim 2.9\cdot 10^{-8}$m, while it remains repulsive for atoms in the free space.
Also, in this range the interaction is reduced by a factor of the order of $10^{-3}$, compared to the free-space case (much more than in the axial-dipoles case).
In the far zone limit, the resonance interaction is strongly suppressed by the waveguide and, analogously to the energy transfer discussed in the previous section, it is virtually zero.

Similar considerations hold for the case of the antisymmetric state, where the resonance interaction is the same that for the symmetric state, except for a change of sign (see Eq. (\ref{equation R})).

All these results clearly show that the waveguide deeply modifies the character and the strength of the  interaction (both in the near- and in the far-zone regimes).  The reduction of the interaction energy is small in the very near zone and it becomes more and more significant when the intermediate region is approached, yielding a complete inhibition of the interaction in the far zone.

\section{Conclusions}
\label{sec:Conclusions}

In this paper we have investigated the energy transfer amplitude between two identical atoms, interacting with the quantum electromagnetic field, placed in a macroscopic environment such as a perfectly conducting cylindrical waveguide. The energy-transfer process studied is also directly related to the resonance interaction energy and force between two identical correlated atoms, prepared in a symmetrical or antisymmetrical Bell-type state.

We have first introduced a general analytical expression for the energy transfer amplitude (and the resonance dipole-dipole interaction between two identical atoms) in a generic structured environment, exploiting the Green's tensor formalism.
We have then considered the specific case of the energy transfer between two identical atoms placed on the axis of a perfectly conducting cylindrical waveguide, which determines a lower cutoff frequency for the electromagnetic modes inside it. We have considered both cases of atomic dipole moments parallel and orthogonal to the axis of the waveguide. When the atomic transition frequency is smaller than the cutoff frequency of the waveguide, we have shown that the presence of the waveguide can significantly change the energy-transfer amplitude, depending on the distance between the two atoms compared to their transition wavelength (near radiationless zone or far radiative zone). We have shown that, when the atomic transition frequency is smaller than the waveguide lower cutoff frequency, the energy transfer process is strongly suppressed in the far zone, while it is only much less influenced in the intermediate and near zone (the latter effect is much larger for radial dipoles than for axial dipoles). A physical interpretation of this result is given. We have also shown that, in this regime, the resonance interaction force between two atoms with radial dipole moments, in the near zone, changes its character from repulsive to attractive. These results show how the presence of the external environment, the cylindrical waveguide in our case, can significantly change atomic radiative processes such as the resonance energy transfer and the resonance dipole-dipole interaction energy, yielding possibility of controlling them through the environment (mainly the waveguide radius, in the present case) and even inhibiting them.

\acknowledgments
The authors gratefully acknowledge financial support from the Julian Schwinger Foundation.

\end{document}